\documentclass[%
 reprint,
nofootinbib,
 amsmath,amssymb,
 aps,
]{revtex4-1}

\usepackage{graphicx}
\usepackage{dcolumn}
\usepackage{bm}
\usepackage{hyperref}

\begin{document}

\title{Non-Existence of Black Holes with Non-Canonical Scalar Fields}
\author{Alexander A. H. Graham}
\email{A.A.H.Graham@damtp.cam.ac.uk}
 \author{Rahul Jha}%
 \email{ R.Jha@damtp.cam.ac.uk}
\affiliation{%
Department of Applied Mathematics and Theoretical Physics\\
Centre for Mathematical Sciences\\
University of Cambridge\\
Wilberforce Road, Cambridge, CB3 0WA, UK 
}%

\date{\today}

\begin{abstract}
We study the existence of stationary black holes with a non-canonical scalar field as a matter source. We prove a simple black hole no-hair theorem which rules out the existence of asymptotically flat black holes which are static or stationary and axisymmetric possessing scalar hair for a wide class of non-canonical scalar field theories. This applies to scalar field theories which are of the form of K-essence theories. In particular, we rule out the existence of such black holes in the ghost condensate model, and in large sectors of the Dirac-Born-Infeld model. 
\begin{description}
\item[PACS numbers] 04.70.Bw, 04.50.Kd
\end{description}
\end{abstract}


\maketitle


\section{\label{sec:level1}Introduction}
One of the most remarkable properties of the Kerr-Newman solution is that it is essentially unique. More precisely, it is widely believed to be the only stationary, asymptotically flat, single black hole solution of the Einstein equations with realistic matter sources\footnote{If multi-black hole solutions are considered there is also the Majumdar-Papapetrou solution \cite{hartle72}.} \cite{wald84, hawking73, chandrasekhar, chrusciel12, bekenstein98}. The set of results which support this conjecture are collectively known under the name of black hole uniqueness theorems, or no-hair theorems in the language of Wheeler. They are probably one of the crowning achievements of the golden age of relativity.

Black hole no-hair theorems generally fall into two classes. The first, which are probably the ones of most physical importance, concern the existence of black hole solutions to the vacuum (or at most electrovac) Einstein equations. Israel was the first to provide such a result by showing that the Schwarzschild solution exhausts the family of static, asymptotically flat, vacuum black holes \cite{israel67}, modulo some technical assumptions which have now been removed \cite{robinson77}. This was quickly generalised to show that the Reissner-Nordstr\"om solution is similarly the unique static, asymptotically flat, electrovac black hole \cite{israel68}. A few years later the proof was generalised to the stationary case. The key step was Hawking's proof \cite{hawking72} that stationary black holes are either static or axisymmetric and have horizons which are topologically spherical, and Carter \cite{carter71} and Robinson's \cite{robinson75} demonstration that the Kerr black hole is the only stationary, axisymmetric, asymptotically flat black hole with spherical horizon topology. This was later generalised to the electrovac case \cite{mazur82}. Although there are still some technical issues regarding analyticity assumptions the theorem can be taken as essentially established in the electrovac case, although it is still an open problem how to extend these results to higher dimensions, to asymptotically (anti-) de Sitter spacetimes and to alternative theories of gravity.

The second line of direction is to extend these results to different matter models. This is arguably of less physical interest, since there is no evidence that any long-range forces exist besides the gravitational and electromagnetic ones. Despite this, it is of obvious theoretical interest to see how far the above results can be extended. Moreover, given that many theories of beyond the standard model physics predict the existence of additional scalar degrees of freedom, investigating the black hole structure of these theories can help us decide on their viability or otherwise. There are really two questions to consider: are there any stationary black holes beyond those found in general relativity, and if there are can they be characterised by a set of global conserved charges for the field? The answer to this question does depend upon the matter chosen, as it is known uniqueness can fail in both of these senses -- in non-Abelian Yang-Mills theories there exist families of black holes with nonzero gauge fields, but which have the same mass and global Yang-Mills charges \cite{bizon90}.

One obvious class of matter models to investigate are scalar fields, and indeed the issue has been considered since the 1970s \cite{chase70}. In 1972 Bekenstein gave a beautiful argument which rules out the existence of stationary, asymptotically flat black holes with scalar hair for canonical scalar fields \cite{bekenstein72a, bekenstein72b, bekenstein72c}. The argument is especially simple since it follows entirely from the scalar field equation of motion. Suppose there does exist a regular, stationary, asymptotically flat black hole solution with a non-trivial scalar field profile. Provided the scalar field obeys the weak energy condition the rigidity theorem holds and the spacetime is either static or stationary and axisymmetric. It must solve the field equation for the scalar field,
\begin{equation} \label{1.1}
\Box\phi-V_{,\phi}(\phi)=0. 
\end{equation}
Now multiply Eq. \eqref{1.1} by \(\phi\) and integrate over a region \(\Sigma\) from outside the black hole horizon to infinity. Integrating by parts on the first term gives the identity
\begin{equation} \label{1.2}
-\int_{\Sigma}(\nabla_{a}\phi\nabla^{a}\phi+\phi{}V_{,\phi}(\phi))\sqrt{-g}d^{4}x+\int_{\partial\Sigma}\phi\nabla^{a}\phi{}dS_{a}=0, 
\end{equation}
where \(dS_{a}=n_{a}d\sigma\) and \(n^{a}\) is the normal of the boundary \(\partial\Sigma\). Now the boundary term is composed of two separate pieces, one at infinity and one at the horizon, and both vanish separately. The contribution at infinity vanishes since the solution is asymptotically flat, which requires \(\phi\) to decay appropriately at infinity; the horizon contribution vanishes because the horizon is a null surface\footnote{More precisely, using the Cauchy-Schwarz inequality we have that \(|\phi\nabla^{a}\phi{}dS_{a}|\leq\sqrt{\phi^{2}\nabla^{a}\phi\nabla_{a}\phi{}dS^{b}dS_{b}}=0\) since the horizon is null and the scalar field is assumed to be regular at the horizon. Notice that the assumption concerning regularity of the scalar field is crucial to the argument. If it is relaxed exact solutions are known to exist; a famous example is the Janis-Newman-Winicour solution \cite{janis68}, which is the general static, spherically symmetric solution of the Einstein equations with a massless scalar field. It describes a naked singularity in general.}. However, since the solution is static or stationary and axisymmetric it can be shown that \(\nabla_{a}\phi\nabla^{a}\phi\) is positive semi-definite outside the horizon. Thereby if \(\phi{}V'(\phi)\) is as well there is no way this identity can be satisfied, except possibly by \(\phi=\mbox{constant}\), and so no such black holes exist. In particular, this immediately rules out black hole solutions with a massless or Klein-Gordon field. An alternative proof can be derived by multiplying Eq. \eqref{1.1} by \(V_{,\phi}\) instead of \(\phi\) and following the same steps as above. This version of the result rules out black holes with scalar hair when \(V_{,\phi\phi}>0\). The results have been strengthened \cite{sudarsky95, bekenstein95, heusler96} to rule out the existence of black holes assuming only that the potential is bounded from below, though at the price of only ruling out static, spherically symmetric solutions, and with a somewhat more complex argument. These results are also of interest in being a key technical step in the proof of the non-existence of stationary black holes in scalar-tensor theories of gravity, since such theories are conformally related to Einstein gravity coupled to a scalar field \cite{hawking72b, sotiriou12}.

Despite the large literature on the subject, there has been very little work regarding whether the no-hair theorems hold when the scalar field action has non-canonical kinetic terms (K-essence theories). Indeed, to our knowledge the only results in this direction are due to Bekenstein \cite{bekenstein95}. Bekenstein showed\footnote{Actually, Bekenstein allowed for several scalar fields and a form of coupling between them.} that there exist no static, spherically symmetric black holes with a scalar hair in the Einstein-K-essence system, provided that the action obeys certain conditions which in our notation translate into the conditions \(P(\phi,X)\leq0\) and \(P(\phi,X)_{,X}\geq0\). Unfortunately, Bekenstein's result has several deficiencies: it does not rule out rotating black holes existing, or indeed static but non-spherically symmetric black holes, and it does not apply to the actions often considered today. The proof also requires use of the Einstein equations. Given the importance of K-essence models in cosmology it seems worthwhile to revise the subject. Here we give a simple proof, modelled on Bekenstein's original argument, which rules out the existence of scalar hair for static or stationary and axisymmetric black holes in a large class of K-essence models. 

Note that our metric signature is \((-,+,+,+)\).

\section{K-essence theories}
The scalar field theories we consider in this paper are defined by the action
\begin{equation} \label{2.1}
S=\int{}d^{4}x\sqrt{-g}P(\phi,X),
\end{equation}  
where \(X=-\frac{1}{2}\nabla_{a}\phi\nabla^{a}\phi\). A canonical scalar field is given by the choice \(P=X-V(\phi)\). The main motivation for these theories comes from cosmology: they were first introduced in Ref. \cite{mukhanov99} as a new model of inflation and later studied as a dark energy model \cite{chiba00, ap00, ap01}. The idea is that by an appropriate choice of \(P(\phi,X)\) one can construct a cosmology which accelerates under the influence of the scalar field, but with the acceleration driven by the kinetic rather than potential energy of the scalar field. They are an alternative to quintessence models -- where acceleration occurs due to the scalar field slow-rolling down its potential -- and are thereby known as K-essence theories. 

The equations of motion for the scalar field are easily found by varying the action with respect to \(\phi\) to yield
\begin{equation} \label{2.2}
\nabla_{a}(P_{,X}\nabla^{a}\phi)+P_{,\phi}=0. 
\end{equation}
The energy-momentum tensor of the scalar field is also easily shown to be 
\begin{equation} \label{2.3}
T_{ab}=P_{,X}\partial_{a}\phi\partial_{b}\phi+Pg_{ab}. 
\end{equation}
Notice that it takes the form of a perfect fluid with pressure \(P_{\phi}=P\), density \(\rho_{\phi}=2XP_{,X}-P\) and 4-velocity \(u_{a}=\partial_{a}\phi/\sqrt{2X}\).

It is important to note that allowing non-canonical kinetic terms into the theory comes at the price of potentially severe instabilities. There are two, related, concerns: the presence of ghosts in the theory and the presence of superluminal modes. The issue has been studied by several authors in the cosmological context \cite{mukhanov99b, hamed04, piazza04, lim05}. They concluded the following. Classical stability of the theory is guaranteed if the sound speed,
\begin{equation} \label{2.4}
c_{s}^{2}=\frac{P_{,X}}{P_{,X}+2XP_{,XX}}, 
\end{equation}
is always positive. This is also the same condition for the theory to admit a well-posed initial value formulation \cite{lim05}, which again is essential for the theory to make sense classically. Quantum stability, which requires that the perturbed Hamiltonian about a background solution is positive, demands that the following conditions be met: 
\begin{equation} \label{2.5}
P_{,X}\geq0,\ {}\ P_{,X}+2XP_{,XX}\geq0,\ \mbox{ and } P_{,\phi\phi}\leq0. 
\end{equation}
It should be noted, though, that it has been argued \cite{piazza04} that the last condition need not be imposed for a physically acceptable theory. Causality is manifestly maintained when the stability conditions \eqref{2.5} hold if \(XP_{,XX}\geq0\) (though see Ref. \cite{babichev08}). Although we have in mind stable examples of K-essence theories we stress that our theorem below requires none of these assumptions.
 
\section{Black hole no-hair theorem}
Let us now give a more precise statement of our black hole no-hair theorem. Consider a asymptotically flat, four-dimensional black hole spacetime which is either static or stationary and axisymmetric, with a minimally coupled scalar field having an action of the form \eqref{2.1}. Moreover, let us assume that the scalar field shares the same symmetries as the spacetime metric. 
Then the only solutions of the Einstein equations with this matter source are of the form $\phi=\mbox{constant}$, provided that 
\begin{equation} \label{3.0}
P_{,X}>0 \mbox{ and } \phi{}P_{,\phi}\leq0 \mbox{ or } P_{,X}<0 \mbox{ and } \phi{}P_{,\phi}\geq0. 
\end{equation}
The last condition covers fields which violate the null energy condition.

We now provide our proof of the no-hair theorem for this class of matter, modelled upon Bekenstein's original argument. The proof follows by contradiction. Assume there does exist a asymptotically flat black hole solution of the Einstein equations which is static or stationary and axisymmetric, with a regular, K-essence scalar field. The scalar field must then satisfy its equation of motion, Eq. \eqref{2.2}. Now multiply Eq. \eqref{2.2} by \(\phi\) and integrate over a region \(\Sigma\) from outside the black hole horizon to infinity. Upon integrating the first term by parts we arrive at the identity
\begin{equation} \label{3.1}
-\int_{\Sigma}(P_{,X}\nabla_{a}\phi\nabla^{a}\phi-\phi{}P_{,\phi})\sqrt{-g}d^{4}x+\int_{\partial\Sigma}\phi{}P_{,X}\nabla^{a}\phi{}dS_{a}=0, 
\end{equation}    
As for the canonical case the boundary is composed of two parts: a contribution from infinity (both spatial and future and past null infinity) and a contribution from the black hole horizon. The contribution from infinity to the boundary term will vanish provided that \(\phi\rightarrow0\) as \(r\rightarrow\infty\) if the solution is asymptotically flat. This follows because use of the equation of the motion of the scalar field shows that all solutions asymptotically obey \(P_{,X_{i}}\nabla^{a}\phi_{i}\sim\frac{1}{r^{2}}\), which suffices to ensure the boundary term vanishes\footnote{The exception is when \(P_{,\phi\phi}(0)\neq0\). In this case the scalar field will in general fall off exponentially fast.}. The horizon contribution vanishes as the scalar field and its first derivative are regular at the horizon (since otherwise the components of energy-momentum tensor would be divergent at the horizon) and since the solution is stationary we may replace spacetime with spatial indices, so the Cauchy-Schwarz identity may be used (as \(g_{ij}\) is positive definite outside the horizon) to show that
\begin{equation} \label{3.2}
\left|\int_{\partial\Sigma}\phi{}P_{,X}\nabla^{a}\phi{}dS_{a}\right|\leq\int_{\partial\Sigma}\sqrt{\phi^{2}P_{,X}^{2}\nabla^{a}\phi\nabla_{a}\phi{}dS_{b}dS^{b}}=0, 
\end{equation}
where we used that the horizon is a null surface (notice that as the spacetime is stationary then \(dS_{0}=0\)) \cite{bekenstein72b, bekenstein72c}. Note that even if the scalar field is not \(C^{\infty}\) then it cannot be the case that any of the terms in the second line of \eqref{3.2} diverge at the horizon, since this would mean the energy-momentum tensor of the scalar field would diverge and it could not be considered a regular solution of the Einstein equations. We are then left with the following integral identity which must be satisfied for solutions to exist:
\begin{equation} \label{3.3}
\int_{\Sigma}(P_{,X}\nabla_{a}\phi\nabla^{a}\phi-\phi{}P_{,\phi})\sqrt{-g}d^{4}x=0. 
\end{equation}
The final step is to show that \(\nabla_{a}\phi\) is spacelike and so \(\nabla_{a}\phi\nabla^{a}\phi\geq0\) outside the horizon. In the static case this follows as we may choose coordinates so that \(g_{0i}=0\), so \(\partial^{0}\phi=0\) and  
\begin{align} \label{3.31}
&g_{ab}\partial^{a}\phi\partial^{a}\phi=g_{ij}\partial^{i}\phi\partial^{j}\phi= \nonumber
\\
&g_{11}(\partial^{1}\phi)^{2}+g_{22}(\partial^{2}\phi)^{2}+g_{22}(\partial^{2}\phi)^{2}\geq0,   
\end{align}
where in the last step the Cotton-Darboux theorem was used to choose the spatial coordinates such that \(g_{ij}\) is diagonal \cite{chandrasekhar}. Hence \(\nabla_{a}\phi\nabla^{a}\phi\) is a sum of squares and positive semi-definite. In the stationary and axisymmetric case we also have that \(\partial^{0}\phi=0\), since coordinates may be chosen so that there is only one off-diagonal component of the metric and the scalar field does not depend upon time  or the coordinate adopted to axisymmetry. This means that Eq. \eqref{3.31} still holds in this case and \(\nabla_{a}\phi\) is spacelike outside the horizon.

Clearly then, if \(P_{,X}\) is of definite sign and \(\phi{}P_{,\phi}\) is of definite, but opposite, sign then Eq. \eqref{3.3} cannot be satisfied, and hence we conclude that no such black hole solutions can exist bar the trivial \(\phi=\mbox{constant}\). Indeed, when there is a potential term in the action the only solution will in general be \(\phi=0\). This completes the proof. In particular, if the action is purely kinetic with \(P_{,X}\) of a fixed sign no such solutions can exist. It should be noted that the proof also assumes implicitly that \(\phi\rightarrow0\) as \(r\rightarrow\infty\), so it does not cover Higgs-like potentials. It does not require explicit use of the Einstein equations though.

Notice that we have not used the rigidity theorem in the proof, since it requires the weak energy condition to be satisfied. If the scalar field model does satisfy the weak energy condition then our theorem combined with the rigidity theorem rules out the existence of stationary, asymptotically flat black holes which obey conditions \eqref{3.0}. Our results do not rule out in principle the existence of stationary, but non-axisymmetric, black holes with scalar fields which violate the null energy condition: it is an interesting question if any in fact exist. Note also that our theorem assumes that the scalar field does not depend upon time, or upon the coordinate adopted to axisymmetry. 

An alternative theorem can be derived on the assumption that \(P_{,X\phi}=0\) even if \(P_{,X}\) and \(P_{,\phi}\) are non-zero. This version in particular applies to Lagrangians of the form \(P=K(X)-V(\phi)\). In this case, multiplying Eq. \eqref{2.2} by \(P_{,\phi}\), instead of \(\phi\), and following the same steps as before yields the following identity which must be satisfied for black holes to exist:
\begin{equation} \label{3.4}
\int_{\Sigma}(P_{,X}P_{,\phi\phi}\nabla_{a}\phi\nabla^{a}\phi-P_{,\phi}^{2})\sqrt{-g}d^{4}x=0. 
\end{equation}
This identity rules out asymptotically flat black holes which are static or stationary and axisymmetric with scalar hair when \(P_{,X}>0\) and \(P_{,\phi\phi}<0\), or when \(P_{,X}<0\) and \(P_{,\phi\phi}>0\). 

This result can be generalised trivially to multiple scalar fields, provided that there is no coupling between the fields. In that case the above argument goes through for each scalar field, and, provided the conditions hold for each, then no black holes with multiple scalar hair can exist. Notice also that our proof does not require use of the Einstein equations and carries over to a large extent in modified theories of gravity, provided the scalar field is minimally coupled. Although our theorem was proved in four dimensions, there is a trivial extension to higher dimensions in the static, spherically symmetric case. This can be seen by noting that in this case the metric can always be put in diagonal form, so \(\nabla_{a}\phi\) is manifestly spacelike outside the horizon.

\section{Examples}
Let us now consider some specific applications of the theorem. Note that outside the horizon of a stationary black hole \(X\leq0\).

\subsection{Ghost condensate model}
The first model we consider is the ghost condensate model. This has the action
\begin{equation} \label{4.1}
P=-X+\frac{X^{2}}{M^{4}}, 
\end{equation}
where \(M\) is a constant. The theory was proposed in Ref. \cite{hamed04}. The idea of this model is that while the first term in the action is ghost-like it can be stabilised by the presence of the higher derivative second term while still admitting self-accelerating solutions. Since \(P_{,X}=-1+2X/M^{4}\leq-1\), the above theorem immediately rules out the existence of static or stationary and axisymmetric black holes with scalar hair in this theory, a result which does not seem to have been noted in the literature before. 

\subsection{Dilatonic ghost condensate model}
The dilatonic ghost condensate model is a fairly straightforward generalisation of the ghost condensate model. It was introduced in Ref. \cite{piazza04}, and is based on the action
\begin{equation} \label{4.2}
P=-X+\frac{X^{2}e^{\lambda\phi}}{M^{4}}, 
\end{equation}
where \(M\) and \(\lambda\) are constants. In this case the integrand in Eq. \eqref{3.3} is easily seen to be
\begin{equation} \label{4.3}
2X-\frac{e^{\lambda\phi}X^{2}}{M^{4}}(4+\lambda\phi),  
\end{equation}
which is negative for \(\lambda>0\) and \(\phi\geq0\) or for \(\lambda<0\) and \(\phi\leq0\). Thereby, no such black hole solutions exist for this theory provided these conditions hold. Alternatively, one can view Eq. \eqref{4.3} as a consistency condition for solutions to exist -- it is required to change sign at least once. 

\subsection{Ghost field}
Consider now a theory similar to a canonical scalar field, except that the scalar field is now ghost-like:
\begin{equation} \label{4.4}
P=-X-V(\phi). 
\end{equation}
We consider this only for completeness, since this model is highly unstable. In this case Eq. \eqref{3.3} now reads
\begin{equation} \label{4.5}
\int{}(2X+\phi{}V_{,\phi})\sqrt{-g}d^{4}x=0. 
\end{equation}
It is clear then that no black holes can exist with a massless ghost field, or when \(\phi{}V_{,\phi}\leq0\). This is the opposite condition to the one derived for canonical scalar fields: it rules out black holes existing with a tachyonic mass (\(V=-\frac{1}{2}m^{2}\phi^{2}\)), but not a regular mass term. For discussion on the existence of phantom black holes and some analytic solutions see Ref. \cite{bronnikov06} and cited literature.

Similar remarks apply equally if we add a potential to the ghost condensate or dilatonic ghost condensate models. In these cases our no-hair result still applies provided that \(\phi{}V_{,\phi}\leq0\). This means in principle there is no reason why black holes might not exist in ghost condensate models if a mass term is added to the action. 

\subsection{DBI model}
The final model we consider is the well known Dirac-Born-Infeld (DBI) model, which was first introduced in Ref. \cite{silverstein04} based on string theory motivations, and has attracted much attention due to its interesting non-Gaussianity predictions \cite{alishahiha04}. Although normally studied as an inflation model it has been considered as a K-essence theory \cite{martin08}. The action of this theory is
\begin{equation} \label{4.6}
P=f(\phi)\left(1-\sqrt{1-\frac{2X}{f(\phi)}}\right)-V(\phi), 
\end{equation}
where \(f(\phi)\) is known as the warp factor. In this case an easy computation shows the integrand in Eq. \eqref{3.3} to be
\begin{eqnarray} \label{4.7}
&&\phi\left[f_{,\phi}\left(\sqrt{1-\frac{2X}{f}}-1\right)+\frac{Xf_{,\phi}}{f}\left(1-\frac{2X}{f}\right)^{-\frac{1}{2}}\right] \nonumber\\&&-2X\left(1-\frac{2X}{f}\right)^{-\frac{1}{2}}+\phi V_{,\phi}.
\end{eqnarray}
When \(\phi{}V_{,\phi}\geq0\), \(\phi{}f_{,\phi}\geq0\) and \(f>0\) the last two terms are manifestly positive, and it can be shown that the first two terms combined are also positive. Hence, no such black hole solutions exist in this model provided these conditions are met. An example where this result applies is the common choice of \(f(\phi)=f_{0}\phi^{4}\). This also holds when a mass term is added to the action. 

\section{Conclusions}
In this paper we have studied the existence of asymptotically flat black hole solutions which are static or stationary and axisymmetric in Einstein gravity coupled to a non-canonical scalar field. We were able to prove a remarkably simple black hole no-hair theorem which rules out the existence of such black holes in a large class of non-canonical scalar field theories. In particular, we ruled out the existence of such black holes in the ghost condensate model, and in the DBI model provided the warp factor and potential obey certain conditions. Our results are also generalizable when potentials are added.

One of the interesting results we found is that while stationary black holes in the ghost condensate theory are ruled out, if we add a mass term to the action there does not seem to be any reason for such black holes not to exist. Indeed, given that such 'phantom' black holes have been found \cite{bronnikov06} we suspect they do exist. We have been unable to solve the field equations for realistic choices of \(V(\phi)\), but it seems a problem worthy of further investigation.

It is important to note that we have only ruled out the existence of stationary solutions: we have said nothing about their existence when the scalar field or geometry is allowed to be time-dependent. In fact, one might consider it a more natural assumption in K-essence theories designed to explain late-time acceleration. For instance, in the ghost condensate model the attractor solution of the Friedmann equations is a de Sitter state with \(\phi\propto{}t\), and so one might expect the scalar field around massive objects to necessarily evolve in time. There have in fact been several studies \cite{frolov04, mukohyama05, babichev06, akhoury09, akhoury11} which have investigated the accretion of a K-essence field onto a black hole (usually assuming the flow is steady state). 

Another interesting direction of research is to investigate to what extent the theorem generalises to higher dimensions. Although the results hold in the static, spherically symmetric case, it is not clear to what extent they generalise when the metric is only stationary. This is because it is unclear that \(\nabla_{a}\phi\) is always spacelike outside the horizon. Similarly, for fields which do not satisfy the weak energy condition then the rigidity theorem does not hold, and so it is possible that there exist stationary, but non-axisymmetric, black holes with scalar hair not covered by our theorem. It would be interesting to resolve whether or not such black holes exist.

We must also stress that not all non-canonical scalar fields have actions of the form of Eq. \eqref{2.1}, in particular the Lagrangian could depend upon the second derivative (or higher) of the scalar field. Usually this will mean that the equations of motion are higher than second order, but in some cases this can be avoided. A good example are Galileons \cite{nicolis09}, in which the equation of motion is second order even though the Lagrangian explicitly depends on the second derivative of \(\phi\). This is achieved through careful choice of the Lagrangian. Although there have been some investigations concerning the existence of black holes in these theories \cite{hui13, sotiriou13, sotiriou14} most of them have only studied the existence of static, spherically symmetric solutions. Our methods could in principle be applied to such theories, provided that the theory possesses no potential, as in this case the scalar field equation of motion can be written as \(\nabla_{a}J^{a}=0\). If we follow the same steps as detailed in section III, with this equation of motion, we can prove a black hole no-hair theorem for these theories provided that \(\nabla_{a}\phi{}J^{a}\) is of definite sign outside the horizon. Unfortunately, it does not seem to apply to most of the theories of interest, since \(J^{a}\) explicitly depends upon the second derivatives of \(\phi\).

Finally, it must also be pointed out that our results only hold assuming the scalar field is uncoupled to other matter. If there is a direct coupling to, say, the electromagnetic field then stationary black holes may carry scalar hair due to the charge. Indeed such solutions are known explicitly for the static, spherically symmetric case -- they were first discovered in the string theory context (see Ref. \cite{garfinkle91} and references therein). Analogous solutions almost certainly exist for K-essence scalar fields, although finding them analytically is likely challenging.

\section*{Acknowledgements}
A.A.H.G. is supported by the STFC. R.J. is supported by the Cambridge Commonwealth Trust and Trinity College, Cambridge. We thank John Barrow and Anne-Christine Davis for helpful discussions and comments on the draft. Special thanks are due to Harvey Reall for spotting an error in an earlier version of the paper.


\end{document}